\documentclass[pre,aps,twocolumn,psfig,epsfig,showpacs,showkeys,floatfix,superscriptaddress]{revtex4-1}
\usepackage{color}
\usepackage{graphicx,amssymb,epsfig,epsf,amsmath,ulem}
\usepackage{soul} 
\usepackage[utf8]{inputenc}

\newcommand{\be}{\begin{equation}}
\newcommand{\ee}{\end{equation}}
\newcommand{\bea}{\begin{eqnarray}}
\newcommand{\eea}{\end{eqnarray}}

\begin{document}
\title{Relaxation of Shannon entropy for trapped interacting bosons with dipolar interactions}

\author{S. Bera}
\affiliation{Department of Physics, Presidency University, 86/1 College Street,
Kolkata 700073, India.}
\author{S. K. Haldar}
\affiliation{Department of Mathematics, University of Haifa, Haifa 3498838, Israel}
\affiliation{Haifa Research Center for Theoretical Physics and Astrophysics, University of Haifa, Haifa 3498838, Israel}
\author{B. Chakrabarti}
\affiliation{Department of Physics, Presidency University, 86/1 College Street,
Kolkata 700073, India.}
\affiliation{Abdus Salam International Center for Theoretical Physics, Strada Costiera 11, 34151 Trieste, Italy.}
\author{A. Trombettoni}
\affiliation{CNR-IOM DEMOCRITOS Simulation Center, Via Bonomea 265, 
I-34136 Trieste, Italy.}
\affiliation{Scuola Internazionale di Studi Avanzati (SISSA) and INFN, Sezione di Trieste, Via Bonomea 265, I-34136 Trieste, Italy.}
\author{V. K. B. Kota}
\affiliation{ Physical Research Laboratory, Navrangpura, Ahmedabad 380009, India}
\date{\today}

\begin{abstract}
  We study the dynamics of dipolar bosons in an external harmonic trap. We monitor the time evolution of the occupation in the natural orbitals and
  normalized first- and second-order Glauber's correlation functions. We focus in particular on the relaxation dynamics of the Shannon entropy. Comparison with the corresponding results for contact interactions is presented. We observe significant effects coming from the presence of the non-local repulsive part of the interaction. The relaxation process is very fast for dipolar bosons with a clear signature of a truly saturated maximum entropy state. We also discuss the connection between the entropy production and the occurrence of correlations and loss of coherence in the system. We identify the long-time relaxed state as a many-body state retaining only diagonal correlations in the first-order correlation function and building up anti-bunching effect in the second-order correlation function.
\end{abstract}
\maketitle
\section{Introduction} \label{intro}
The investigation of non-equilibrium quantum dynamics has
been stimulated by the remarkable progress in experimental techniques.
Ultracold atomic gases and trapped ions are a
test-bed for such studies as they offer a very good
isolation from the environment~\cite{kauf,natphys:13,chen,nat440:14,nat419:11,nat465:12,lan,lan1,science337}. 
A forefront research in this direction aims at characterizing the
dynamical properties of isolated quantum many-body systems
~\cite{science337,santos85,santosprl108,pla350,ming,calabrese}.
In this context, the onset of thermalization
in isolated quantum systems caused by the interparticle interaction
has received great interest~\cite{pre2,pre4,pre5,pre12a,pre12b,physreport}.
The necessary condition for the thermalization is the statistical
relaxation of the isolated system to equilibrium.
In some recent
studies~\cite{pre40,pre41,pre44,pre42,pre43,pre49,pre46,pre47,pre45,pre48}
the viability of thermalization has been associated with the onset of
quantum chaos. The latter in systems of interacting Fermi or Bose particles
implies a pseudorandomness depending mainly on the strength
of the interparticle interaction. In the recent study 
of interacting spin $\frac{1}{2}$ system,
the delocalization of the eigenstates in the energy shell approach was
investigated~\cite{santos85,santosprl108}. The eigenstate
thermalization hypothesis (ETH)~\cite{pre1}
predicts that the expectation value of few-body observable
should correspond to the prediction of microcanonical
ensemble~\cite{pre11,pre8,pre10}. Although a vast amount of works
exist to characterize the delocalized eigenstates and
its connection with the statistical relaxation, many
open questions which still are studied :\\
 a) How and in which timescale the system relaxes? \\
 b) How the time evolution of the entropy and the
 onset of statistical relaxation are connected? \\
 c) What is the link between the production of entropy and
 both the build up of correlations and the loss of coherence? \\
 d) What is the effect of long-range interactions? 

The last question {\it d)} is motivated by the fact that many investigations
have focused on contact/short-range investigations. In this respect
it would be interesting to compare results for long-range interactions with short-range findings
and with recent results for quantum systems with long-range
couplings~\cite{vodola14,gong16,lepori16,celardo16,defenu16,lepori17,defenu17,igloi18,blass18,defenu18,lerose19}.
Ultracold atoms with dipole-dipole interactions are a popular setup to investigate the effect of non-local
interactions~\cite{baranov12}. Dipolar atoms in quasi-one-dimensional traps can be experimentally realized
and provide a tool to explore a rich many-body physics~\cite{1d_3d_1,1d_3d_2}. It is important to point out that
one can realize also several coupled one-dimensional systems, with a tunable couplings~\cite{tang19}.

In this paper, we focus on the role of dipolar interaction in the relaxation dynamics of interacting bosons in a
1D harmonic oscillator (HO) trap. The comparison between the short-range contact interactions and long-range dipolar ones
is also presented. Ultracold dilute Bose gases are very often well described by the contact interaction,
defined as
\begin{equation}
  \hat{V}(x_{i}-x_{j}) = \lambda \delta (x_{i}-x_{j}),
  \label{cont}
\end{equation}
where the dimensionless parameter $\lambda$ is the strength of the 1D contact interaction~\cite{olshanii}.
For dipolar interaction, in quasi-1D geometries, the effective non-local two-body interaction term $V$ can be obtained by integrating
over the transverse directions~\cite{sinha}. Here we consider dipolar interaction of the form
\begin{equation}
  \hat{V}(x_i-x_j)= \frac{g_{d}}{\vert x_i-x_j \vert ^{3} +\alpha },
  \label{dipolar}
\end{equation}
where the dimensionless parameter $g_{d}$ is the strength of interaction and
$\alpha$ is the short-scale cut-off to regularize the divergence at $x_{i}=x_{j}$.
We solve the $N$ body Schr\"odinger equation with very high level of accuracy using the
multiconfigurational time-dependent Hartree method for bosons (MCTDHB)~\cite{ofir77,mctdhb_prl100,sakmann,Axel,fischer15}, with the
many-body ansatz being the sum of the different configurations of $N$ particles distributed over $M$ orbitals.

In our present setup, we consider $N$ dipolar bosons in 1D HO trap. We are going to consider few particles, such as $N=4$, even though
we checked that the obtained results are consistent with findings for higher number of particles, such $N \sim 6$.
Our procedure corresponds to have the non-interacting system and then at time $t=0$ perform a quantum quench from
$g_d=0$ to a finite, possibly large, value of $g_d$. Quantum quenches of the 1D Bose gas with contact
interactions have been deeply investigated in the
literature~\cite{iyer12,kormos13,kormos14,mazza14,sotiriadis14,franchini15,collura16,piroli16,sotiriadis16,franchini16,alba18,kormos18}.

We are going to obtain the results obtained with the dipolar interactions to those obtained by quenching the contact interaction parameter
$\lambda$ from zero to a finite value.
The relaxation is studied by analyzing the time evolution of Shannon information entropy and as well the
normalized first- and second-order Glauber's correlation functions. The contrast
between contact and dipolar interactions is demonstrated by the timescale of relaxation process.
We observe interesting many-body properties in the time evolution of first- and second-order correlation functions in the case
of dipolar interactions. We also demonstrate that the observed 
relaxation
is associated with the loss of coherence in the first-order correlation function and the occurrence of an {\it anti-bunching} effect
in the second-order correlation function. The effect of long-range repulsive tail of the dipolar interaction
makes the dynamics very interesting. The corresponding entropy evolution shows sharp linear increasement
at very short time and then saturation. The first-order correlation is quickly lost as well as a clear anti-bunching effect
is quickly developed for the dipolar interaction.

Since the high resolution image technique allows to probe the spatial correlation functions~\cite{bakr,hung},
the observations made in the present manuscript could be verified in future experiments. 

The paper is organized as follows. In Section~\ref{method}, we give a brief
description of Hamiltonian and the used numerical method. In Section~\ref{key_measures},
we introduce the key quantities that are subsequently analyzed.
In Section~\ref{stat_relax} we presents our results for the post-quench dynamics.
Section~\ref{conclusion} provides a summary and discussion of our results.

\section{The model}\label{method}
The time-dependent Schr\"odinger equation (TDSE) for $N$ interacting bosons is given by 
\begin{equation}
   i\partial_t \vert \Psi \rangle = \hat{H} \vert \Psi \rangle 
\label{TDSE}
\end{equation}
(with $\hbar=1$). The Hamiltonian $\hat H$ is 
\begin{equation}
\hat H(x_{1},x_{2},...,x_{N})=\sum_{i=1}^{N}{\hat {h}(x_{i})} + \Theta (t) \sum_{i<j=1}^{N}\hat{V}(x_{i}-x_{j}).  
\end{equation}
$\hat h(x)$ is the one-body Hamiltonian containing the external trapping potential and the kinetic energy.
We set the external trap as harmonic oscillator trap $[V_{ext}(x_{i})=\frac{1}{2}x_{i}^{2}]$. $V(x_i-x_j)$ is the
two-body interaction which is chosen either as dipolar or - for comparison - contact,
respectively Eq.~(\ref{dipolar}) and
Eq.~(\ref{cont}). The Hamiltonian is in dimensionless units - obtained by dividing the dimensionful Hamiltonian by $\frac{{\hbar}^2}{mL^2}$ ($m$ is the mass of the bosons,
$L$ is an harmonic oscillator length).
$\Theta (t)$ is the Heaviside step function of time $t$, so to have an interaction quench, the dipolar interaction
being abruptly turned on at $t=0$.

To solve the TDSE~(\ref{TDSE}) we expand the many-body wave function in a complete set of time-dependent
permanents with $M$ orbitals. Thus the ansatz for the many-body wavefunction in the MCTDHB approach is 
\begin{equation}
\vert \Psi(t)\rangle = \sum_{\bar{n}}^{} C_{\bar{n}}(t)\vert \bar{n};t\rangle,
\label{many_body_wf}
\end{equation}
where $\vert \bar{n};t\rangle$ are the permanents which are the symmetrized bosonic states
considering all configurations of $N$ particles in $M$ orbitals. The sum in Eq.~(\ref{many_body_wf}) runs on
all possible configurations determined as $N_{conf}$= $\left(\begin{array}{c} N+M-1 \\ N \end{array}\right)$.
In the second quantized representation the permanents are given as 
\begin{equation}
\vert \bar{n};t\rangle = \vert n_{1},...n_{M};t\rangle = \prod_{i=1}^{M}
\left( \frac{ \left( b_{i}^{\dagger}(t) \right)^{n_{i}} } {\sqrt{n_{i}!}} \right) \vert vac \rangle,
\label{2nd_quantized_wf}
\end{equation} 
where $ b_{k}^{\dagger}(t)$ is the bosonic creation operator creating a boson in the time-dependent single particle state
called orbital $\phi_{k}(x,t)$. $\bar{n}=(n_{1},n_{2},...n_{M})$ represent the occupations in the orbitals and
preserve the total number of particles $n_{1}+n_{2}+...+n_{M}=N$. It is to be emphasized
that in the ansatz of the many-body wave-function, both the expansion
coefficients $\{C_{\bar{n}}(t);\sum_{i}^{}n_{i}=N \}$ and the orbitals $\{ \phi_{i}(x,t) \}_{i=1}^{M}$ that build up
the permanents are time-dependent and variationally optimized quantities.
The efficiency of the MCTDHB method comes from the variationally optimized and time-adaptive basis~\cite{sakmann,Axel,rhombik,cao1,cao2}.
In the computation, we limit the size of the Hilbert space. However as the evolution follows from the
time-dependent variational principle~\cite{TDVP}, the error resulting from the truncation of the Hilbert space is minimized by the
basis at any given time. 
In the Section~\ref{stat_relax} we show that with the considered number of particles
several natural orbitals may have a significant and comparable occupation.

\section{Shannon entropy and correlation functions} \label{key_measures} 
The Shannon information entropy (SIE) of the one-body density
in position space is defined as $ S_x(t) = - \int dx \rho(x,t) ln [\rho(x,t)] $ and similarly in the momentum space
as $S_k(t) = - \int dk \rho(k,t) ln [\rho(k,t)]$ where $\rho(x)$ and $\rho(k)$ are the density of the system
in coordinate space and in momentum space respectively. The two SIEs as defined above are two independent
key quantities in the calculation of quantum information in many-body system, as they measure
the delocalization of the corresponding distributions. However, since SIE is based on the one-body density,
one can not infer the presence of correlation in the many-body state. 
We therefore provide an alternative definition of the SIE calculated from the time-dependent coefficients as 
\begin{equation}
S^{info}(t) = -\sum_{\bar{n}}^{}{\vert C_{\bar{n}}(t)\vert}^{2}\ln {\vert C_{\bar{n}}(t)\vert}^{2}.
\label{1BS_c}
\end{equation} 
The many-body measure of the information entropy can be explicitly made by writing
the coefficients as an expectation value in terms of $M$ creation and annihilation operators~\cite{pra2015}.

We observe that in the Gross-Pitaesvkii~\cite{GPbook} and multiorbital mean-field theory~\cite{TDMF},
since just a single coefficient contributes, $S^{info}(t)$ is always zero. Thus the information entropy
measure using MCTDHB basis qualifies when and how well or not a given many-body state can
be captured by mean-field theory. We are also interested in the calculation of normalized first- and
second-order Glauber's correlation functions at many-body level. They indeed may provide additional tools
to study the pathway of the occupation of the different natural orbitals.
We will see how these correlation functions effectively describe the relaxation of the system in terms
of the loss of first-order coherence and emergence of anti-bunching effect in the second-order coherence.

The normalized $p$-th order correlation function is defined as
\begin{eqnarray}
 g^{(p)}(x_1^{\prime},...,x_p^{\prime},x_1, ...,x_p;t)= \\
\frac{\rho^{(p)}(x_1,...,x_p \vert x_1^{\prime},...,x_p^{\prime};t)}
{\sqrt{\prod_{i=1}^p\rho^{(1)}(x_i \vert x_i;t )\rho^{(1)}(x_i^{\prime} \vert x_i^{\prime};t )}}.  \nonumber  
\end{eqnarray}
It is the key quantity to define the spatial $p$-th order coherence.
Here, $\rho^{(p)}(x_1,...,x_p\vert x_1^{\prime},...,x_p^{\prime}; t)$ is the $p$-th order reduced density matrix of the state
$\vert \Psi \rangle$~\cite{RJG}.
In the case of $\vert g^{(p)}(x_1...,x_p,x_1 ...,x_p;t) \vert > 1$ ($<1$), the detection probabilities
of $p$ particles at positions $x_1,...,x_p$ are referred to as (anti-)correlated.
Although it is possible to define the $p$-th order correlation function in momentum space and
one can define the detection probabilities, however for the present work we report results
only for the spatial coherence. The normalized first-order coherence is directly related to
the fringe visibility in interference experiments and it is defined as
 \begin{equation}
 g^{(1)}(x_1^{\prime},x_1;t)= 
\frac{\rho^{(1)}(x_1^{\prime}\vert x_1;t)}
{\sqrt{\rho(x_{1}^{\prime};t) \rho(x_{1};t)}}.
\label{1st_corr}  
\end{equation}
 $g^{(1)}(x_1^{\prime},x_1;t)<1$ means the visibility of interference fringes in the experiment is less than $100 \%$,
 which is referred to as loss of coherence. At variance, $g^{(1)}(x_1^{\prime},x_1;t)=1$ corresponds to
 maximal fringe visibility and is referred to as full coherence.
 
The corresponding second-order correlation function $g^{(2)}(x_{1},x_{2};t)$ is calculated as 
\begin{equation}
 g^{(2)}(x_{1},x_{2};t)= \frac{\rho^{(2)}(x_1,x_2;t)}
{\rho(x_{1};t) \rho(x_{2};t)},
\label{2nd_corr}
\end{equation}
where $\rho^{(2)}$ is the diagonal part of the two-body reduced density matrix. When $g^{(2)}<1$, we refer to it as the
anti-bunching effect, while the case $g^{(2)}>1$ is termed as bunching. $g^{(2)}=1$ signifies that the measures of
two particles at positions $x_{1}$ and $x_{2}$ are stochastically independent.

\section{Results}\label{stat_relax}
To compare the results of dipolar interaction with the contact one,
we fix the interaction strength $g_{d}$ and $\lambda$ requiring the effective interaction
\begin{equation}
  \int \frac{1} {x^3 +\alpha} dx = \int \delta(x) dx = 1,
  \label{integ}
\end{equation}
so that the integral of $\hat{V}$ is the same with $\lambda=g_d$ in our units. Eq. (\ref{integ}) fixes as well
the cut-off parameter $\alpha $. Throughout the computation we keep a number $M$ of orbitals which are sufficient to get
convergence in the measured quantities (for $N=4$ it is enough $M=12$).
The convergence is further guaranteed when the occupation in the last orbital is negligible.
We prepared the ground state of the non-interacting Hamiltonian using the R-MCTDHB package~\cite{Axel,ultracold,Axel1,Axel2}.
The interactions are then abruptly turned on at $t=0$. 

\begin{figure}
\begin{center}
{\includegraphics[width=0.25\textwidth,angle=270]{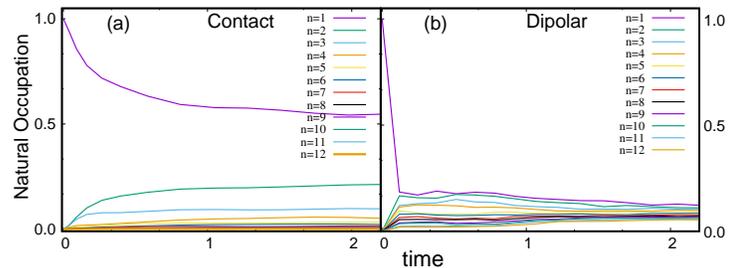}}
\end{center}
\caption{Eigenvalues of the reduced density matrix (i.e., the natural occupations) as a function of time.
  Panel(a): Contact interaction. Panel(b): Dipolar interaction. All quantities shown are dimensionless and $g_d=\lambda=5$
  (the same values of $g_d$ and $\lambda$ are used for the next figures).} 
\label{Fig1} 
\end{figure}

In Fig.~\ref{Fig1} we plot the natural occupations $n_{i}$ as a function of time. The left panel is for contact interaction
and right panel is for dipolar interaction. We choose $g_d=\lambda=5$, and we found qualitatively
very similar results for other values of $\lambda$ we tried. 
Initially at $t=0$,
only the first natural orbital contributes. When the time goes on, the occupation of the higher orbitals start to contribute.
However we get a clear difference with time between the contact and the dipolar interactions. For contact interaction,
mostly one natural occupation, $n_{1}$, dominates throughout the time evolution and other occupations remain comparatively small.
At variance, for dipolar interaction, again initially at time $t=0$ only $n_{1}$ contributes, but at
short time $n_1$ sharply decreases and other orbitals start to populate.
For the rest of evolution we find that several other orbitals almost equally contribute as $n_{1}$ (we checked that the obtained
results do not depend on the chosen value of $M$). 
However, with the considered number of particles
our present computation is unable to present the full-blown $N$-fold occupation of natural orbitals
which can be achieved when the system is quenched to very large values of $g_{d}$, that correspond to crystal-like
states~\cite{budha_order,bera18}.  

\begin{figure}[!]
{\includegraphics[height=0.55\textwidth,angle=-90]{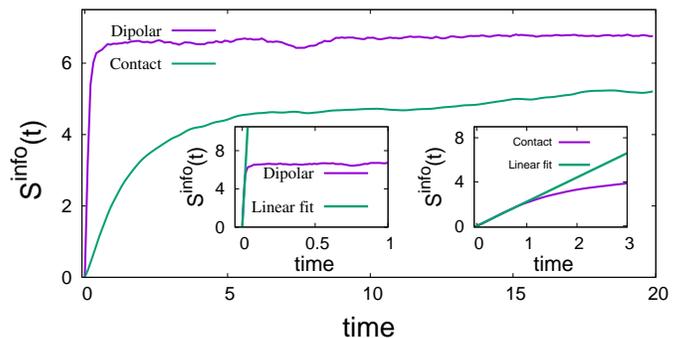}}
\caption{Dynamics of the many-body SIE $S^{info}(t)$. The statistical relaxation is seen by the convergence
  of $S^{info}(t)$ to the maximum entropy value. One has $\Gamma_{c}=0.34$ and $\Gamma_{d}=43.2$.
  The insets present the sharp linear increase fitted with the analytical formula (see text) for small times.}
\label{Fig2} 
\end{figure}

Fig.~\ref{Fig2} reports the dynamics of many-body SIE for dipolar as well as for contact interactions.
The statistical relaxation is manifested by the long-time dynamics when $S^{info}(t)$ saturates to a maximum entropy state.
At very short time we observe linear increase in $S^{info}(t)$ fitted with the analytical formula $S=\Gamma t\ln{P}$,
where $\Gamma$ is determined by the decay probability to stay in the initial ground state and $P$ is the number of
many-body states~\cite{flam}. However the difference in $S^{info}(t)$ for contact and
dipolar interactions can be identified from the corresponding $\Gamma$ values.
For contact interaction the linear increase is determined with a $\Gamma_{c}$
significantly smaller than the corresponding $\Gamma_{d}$ for dipolar interaction.
The very sharp linear increase in the information entropy for dipolar interaction implies the number of principal
components participating in the many-body dynamics increases exponentially very fast -- exhibiting very quick
relaxation process. At long times, the system relaxes to the maximum entropy state. For contact interaction,
$S^{info}(t)$ also has a tendency to achieve a saturation value or maximum entropy value, however the process
of relaxation is very slow which is also quantified by the small value of $\Gamma_{c}$. The
saturation value or maximum entropy value achieved by the system for contact interaction is smaller than that for the dipolar interaction. 

\begin{figure}[!]
\begin{center}
{\includegraphics[width=0.5\textwidth,angle=0]{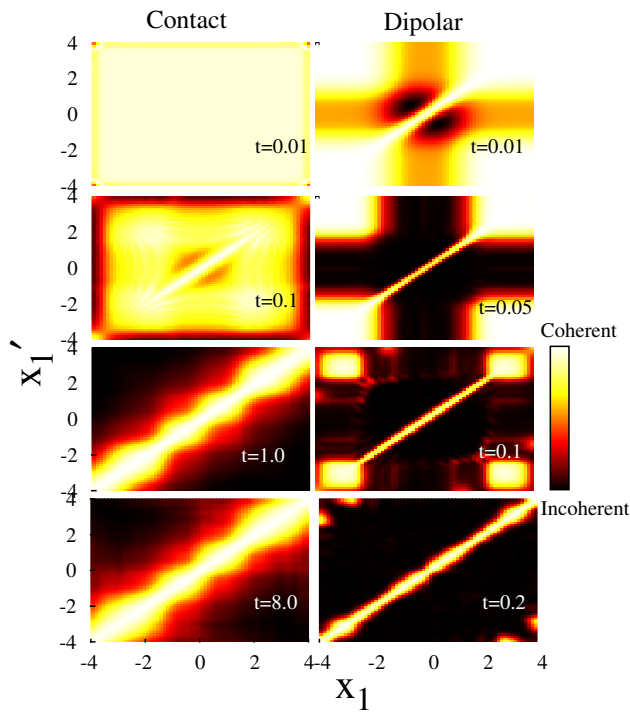}}
\end{center}
\caption{Coherence in the quench dynamics measured with first-order correlation function
  $\vert g^{(1)}(x_{1},x_{1}^{\prime};t) \vert^{2}$. The left column depicts $\vert g^{(1)}\vert^{2}$ for contact interaction
  for the times $t=0.01,0.1,1.0$ and $8.0$ respectively. The right column shows the same for dipolar interaction for time
  $t=0.01,0.05,0.1$ and $0.2$ respectively. For the dipolar interaction $\vert g^{(1)}\vert^{2}$
  becomes close to zero almost everywhere except the diagonal at very short time, in good agreement with the
  quick and large production of the many-body SIE.}
\label{Fig3} 
\end{figure}

Fig.~\ref{Fig3} presents the time evolution of the first-order correlation function
$\vert g^{(1)}(x_{1},x_{1}^{\prime};t) \vert^{2}$ as a function of its two spatial variables $x_{1} $ and $x_{1}^{\prime}$
for various time $t$ and fixed interaction strength quench. For contact interaction,
$\vert g^{(1)}(x_{1},x_{1}^{\prime};t) \vert^{2}$ remains close to unity for all $(x_{1},x_{1}^{\prime})$ for a comparatively long time.
This implies that the system remains coherent. At longer time the off-diagonal correlation
is gradually lost and finally at time $t=1.0$ only the diagonal correlation is maintained. The strong interparticle
repulsion leads to the loss of coherence which is further maintained at larger timescale when the system reaches
to its relaxed state. It is also in good agreement with the relaxation process when $S^{info}(t)$ saturates to maximum entropy
state. In contrast, for dipolar interaction we observe very quick loss of off-diagonal correlation.
$\vert g^{(1)}(x_{1},x_{1}^{\prime};t) \vert^{2}$ is close to unity almost exclusively for $x_{1}=x_{1}^{\prime}$,
away from the diagonal ($x_{1} \neq x_{1}^{\prime}$) is close to zero. This is the effect of
long-range repulsive tail of the dipolar interaction. Thus the relaxed state can be described as a many-body state with many
natural orbitals occupied, maximum entropy and persistence of diagonal first-order correlation. 

\begin{figure}[!]
\begin{center}
{\includegraphics[width=0.5\textwidth,angle=0]{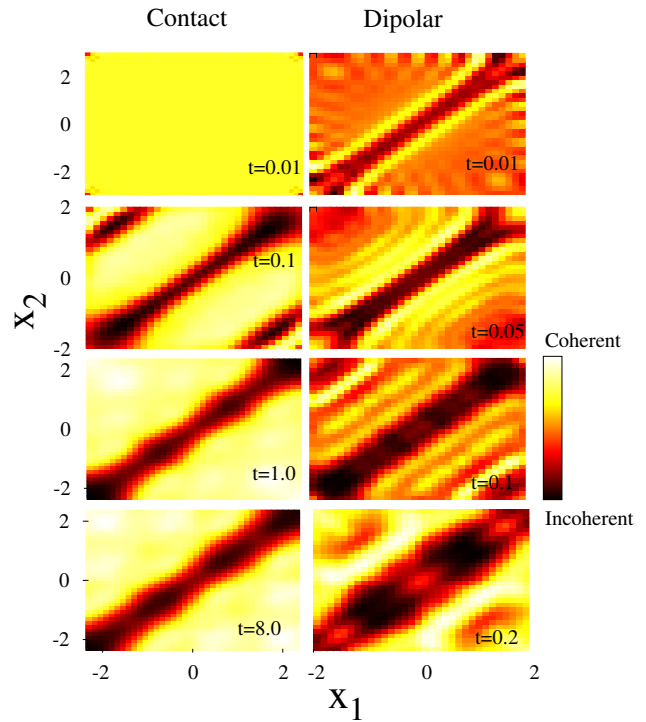}}
\end{center}
\caption{Coherence in the quench dynamics measured with second-order correlation function $g^{(2)}(x_{1},x_{2};t)$.
  The left column depicts $g^{(2)}$ for contact interaction for time $t=0.01,0.1,1.0$ and $8.0$ respectively. T
  he right column shows the same for dipolar interaction for time $t=0.01,0.05,0.1$ and $0.2$ respectively. For dipolar
  interaction the diagonal part of $g^{2}$ is extinguished quickly due to the repulsive long-range tail and the
  anti-bunching effect develops in correspondence of the behaviour of the SIE.}
\label{Fig4} 
\end{figure}

In Fig.~\ref{Fig4}, we plot the corresponding two-body correlation function $g^{(2)}(x_{1},x_{2};t)$
for the same parameters as reported in Fig.~\ref{Fig3}. At small time 
for almost all $(x_{1},x_{2})$, the second-order coherence is maintained for contact interaction.
Whereas the diagonal coherence starts to deplete for dipolar interaction. For larger times (such $t=1$),
$g^{(2)}(x_{1},x_{2};t) \approx 1$ at the off-diagonal $(x_{1} \neq x_{2})$ for contact interaction, whereas the diagonal
is almost vanishing. It means that there is a finite probability of detecting two particles for all ($x_{1}, x_{2}$),
except for the narrow band around the diagonal. The vanishing diagonal part of $g^{(2)}(x_{1},x_{2};t)$ is corresponding
to the anti-bunching effect, as the probability of finding a double occupation along the diagonal is almost zero.
Complete vanishing of the diagonal coherence is maintained at larger times. For dipolar interaction, the
anti-bunching effect appears at very short times such $t \approx 0.01$. With further increasement in time,
the anti-bunching band spreads.
The quick development of the anti-bunching effect for dipolar interaction is also in good agreement with  
our previous observation~\cite{pra2015} when $S^{info}(t)$ quickly attains the saturation value.

We conclude that by observing very quick loss of first-order coherence and the setup of the anti-bunching
effect in second-order coherence may be considered as characterizing the many-body state with maximum entropy.
Since with further increase in time we do not observe any change in entropy production,
first- and second-order correlation functions, we may define the state as a relaxed state.
As the different orders of coherence can be measured experimentally the above relation between the entropy production
and coherence can be directly verified in the experiment to test the process of statistical relaxation.

Our present calculation is done for a finite system of few particles. The natural question is
to verify our present observation for larger bosonic systems. In our previous work~\cite{pra2015}, we have already reported the quench dynamics with contact interaction for a larger number of bosons, however only the first-order correlation dynamics
and its link with the production of entropy have been discussed. In the context of our present computation
we redid the simulation for $N=10$ bosons with contact interaction and observed the similar physics
in the second-order correlation dynamics as observed for $N=4$ bosons. However we are unable
to extend our simulation for $N=10$ bosons with dipolar interaction due to serious convergence problem.
Increasing the particle number the size of the Hilbert space rapidly increases (for $N=4$ bosons distributed over $M=12$ orbitals
the size of the Hilbert space is $N_{conf}=1365$ and the same for $N=10$ bosons is $352716$). However for $10$ bosons,
with dipolar interaction $M=12$ orbitals are not sufficient to achieve convergence. So we checked 
our present observations for $N=6$ bosons distributed in $M=15$ orbitals and the general conclusion drawn for $N=4$ dipolar bosons
remain unchanged.     

\section{Conclusion}\label{conclusion}
In this paper, we considered few dipolar bosons in a 1D harmonic trap. We studied
the statistical relaxation of dipolar bosons and its comparison with that for contact interaction.
We solved the quantum many-body dynamics by MCTDHB and the relaxation is presented through the time evolution of
natural occupation, entropy production, normalized first- and second-order Glauber's correlation functions.
In the long time dynamics, the system relaxes to its maximum entropy state. The effect of
the long-range repulsive tail of dipolar interaction in the dynamics is clearly visible.
The relaxation process for dipolar interaction is quicker than that for contact interaction.
We also presented a link between the production of entropy and the first- and second-order coherences. We observed
that at the time when the many-body system occupied many natural orbitals, then
at the same time the off-diagonal coherence in first-order correlation function is completely lost, and
the anti-bunching effect is exhibited in the two-body correlation.
Thus in our present work we redefine the relaxed state as the many-body state with maximum entropy retaining only the diagonal correlation in $g^{(1)}$ and developing the anti-bunching effect in $g^{(2)}$.
Two remaining and natural open questions are to study the connection of these results with the behaviour of collective modes
and to consider the broader class of long-range interactions with smaller power of interaction,
especially when its value is smaller than the dimension of the considered system.\\

{\em Acknowledgments:}
S. Bera wants to acknowledge Department of Science and Technology (Government of India)
for the financial support through INSPIRE fellowship [2015/IF150245]. B. Chakrabarti acknowledges ICTP
support where the major amount of work has been done. 

\end{document}